\documentclass[amsmath,amssymb,aps,pra,superscriptaddress,10pt,twocolumn]{revtex4-1}

\usepackage{dcolumn}
\usepackage{graphicx}
\usepackage{color}

\begin{document}

\title{Associative detachment of rubidium hydroxide}

\author{Jason N. Byrd}
\email{byrdja@chem.ufl.edu}
\affiliation{Department of Physics, University of Connecticut, Storrs, CT 06269}
\affiliation{Quantum Theory Project, University of Florida, Gainesville, Florida 32611}

\author{H. Harvey Michels}
\affiliation{Department of Physics, University of Connecticut, Storrs, CT 06269}

\author{John A. Montgomery, Jr.}
\affiliation{Department of Physics, University of Connecticut, Storrs, CT 06269}

\author{Robin C\^{o}t\'{e}}
\affiliation{Department of Physics, University of Connecticut, Storrs, CT 06269}

\begin{abstract}
{ We performed calculations of the optimized structure, harmonic
vibrational frequencies and dissociation energies of RbOH and its anion, and
investigate the interactions between Rb and OH$^-$ 
leading to possible associative detachment pathways}.  The
electron affinity of RbOH was computed to be 0.2890 eV, with { a  bond
energy of Rb+OH$^-$ half that of Rb+OH.}  To determine other possible charge loss
pathways, the Rb+OH and Rb+OH$^-$ dissociation curves were computed using couple
cluster methods along all possible collisional angles.  An adiabatic curve
crossing between the neutral and charged molecule was found at the inner wall of
the molecular potential curve for linear geometries.  Associative detachment
rates were estimated using the Langevin ion capture cross-section for
hydroxide.  We find for $v\ge 2$ an associative detachment rate of $>2\times
10^{-9}$ cm$^3$s$^{-1}$, while for $v=0$ and $1$ no appreciable rate exists.
This strong dependence on vibrational level suggests the ability to control the
{ associative detachment rate}  directly.
\end{abstract}

\maketitle

\section{Introduction}

Advances in the formation of cold molecules have opened up avenues into many
branches of the physical sciences \cite{carr2009,dulieu2011}.  For chemical
physics, applications range from precision spectroscopy \cite{demille2008}, to
the study \cite{miranda2011,sawyer2011} and control \cite{quemener2012} of cold
chemical reactions.  Other areas of physics benefit greatly from the study of
cold molecules, such as condensed matter physics \cite{micheli2006}, and the
search for novel quantum gases \cite{santos2000} and phases \cite{recati2003}.
Oxides such as PbO \cite{demille2000}, YO \cite{hummon2013} and TiO
\cite{stuhl2008} have been of interest recently for electron dipole moment
measurements, direct cooling and trapping of molecules respectively, as has the
OH radical \cite{stuhl2013}.  Molecular ions have the advantage of being easily
trapped and cooled using radio frequency traps and sympathetic cooling
\cite{schiller2009} while not involving an oxide radical.  Recent progress on
the oxide ion front has seen the co-trapping \cite{deiglmayr2012} of cold
rubidium and hydroxide.  In this work we investigate the interactions of Rb+OH,
and find that it is possible to control the associative dissociation process of
forming neutral rubidium hydroxyl by accessable laser controlled transitions.


\section{Electronic Structure Calculations}

Electronic structure calculations were performed on RbH, RbO, OH, RbOH and
associated anions using a combination of perturbation and coupled cluster theory
\cite{pople1987}.  Second order M{\o}ller-Plesset (MP2) perturbation theory and
coupled cluster theory with all singles, doubles, and perturbative triples
(CCSD(T)) energy and gradient calculations in this work were carried out using
the CFOUR \cite{CFOUR} and MOLPRO 2010.1 \cite{molpro10_short} quantum chemistry
packages.  Higher order calculations involving CCSDT and CCSDT(Q) (all triples
and perturbative quadruples respectively) were done using the MRCC program of M.
K\'allay \cite{kallay2001}.  For open shell systems, the spin restricted
variants of these theories were used.  Due to the size of rubidium, there are a
number of correlation space choices available for consideration. 
We have adopted the same notation as Sullivan {\it et al.}\cite{sullivan2003}
where valence only calculations (H:1s; O:2s2p; Rb:5s) are referred to as relaxed
valence (rv).  Increasing the correlation space size to involve the first set of
sub-valence orbitals (H:1s; O:1s2s2p; Rb:4s4p5s) results in the relaxed
inner-valence (riv), while spaces including yet deeper orbitals (H:1s; O:1s2s2p;
Rb:3s3p3d4s4p5s) are called riiv and so forth.  Valence (rv) only calculations
involving rubidium and oxygen require extra care, as the usual method of energy
sorting orbitals in selecting the frozen core will fail since the energy of $2s$
orbital of oxygen is below the $4p$ orbital of rubidium.  Failing to properly
choose the core orbitals for valence calculations will lead to significant
errors. 

\begin{table}[b]
\begin{ruledtabular}
\begin{tabular}{llllll}
         & r(M-H) & r(M-O) &
ZPE\footnote{ZPE$\simeq\omega_e/2$.}\cite{abrams1994} & EA & AE\\
\hline
OH                       & 0.9698 &        & 8.53 &        & 162.91\\
Exp.                     & 0.9696\cite{amano1984} 
                                  &        & 8.51\cite{abrams1994} 
                                                  &        & 161.53\cite{huberherzberg}\\
OH$^-$                   & 0.9643 &        & 8.55 & 1.8405 & 176.51\\
Exp.                     & 0.9643\cite{rosenbaum1986}
                                  &        & 8.51\cite{rosenbaum1986}  
                                                  & 1.8277\cite{smith1997}
                                                           & 175.65\cite{martin2001}\\
RbO                      &        & 2.3548 & 0.79 &        & 102.21\\
Exp.                     &        & 2.2541\cite{yamada1999} 
                                           & 0.84\cite{yamada1999} \\
RbO$^-$                  &        & 2.2564 & 0.65 & 0.1002 & 58.90\\
RbH                      & 2.3919 &        & 1.12 &        & 61.39\\
Exp.                     & 2.37\cite{kato1985} 
                                  &        & 1.34\cite{kato1985}\\
RbH$^-$                  & 2.5415 &        & 1.66 & 0.3604 & 47.11\\
RbOH                     & 0.9551 & 2.3408 &11.72 &        & 291.26\\
Theory                   & 0.959\cite{lee2003} 
                                  & 2.472\cite{lee2003} 
                                           &11.36\cite{lee2003} \\
RbOH$^-$                 & 0.9567 & 2.4166 &11.27 & 0.2912 & 247.98
\end{tabular}
\end{ruledtabular}
\caption{\label{geomtable}Computed bond lengths, harmonic zero point energies, electron
affinities (EA) and atomization energies (AE) for RbOH, its constitutes
and their anions (Units are in angstroms, electron volts and $10^{-3}$ a.u. as
appropriate).}
\end{table}

While there are many basis sets available for the first row elements, the basis
set selection for rubidium is sparse.  This is further complicated by the need
for diffuse functions to accurately describe electron affinities
\cite{kendall1992}.  Previous calculations \cite{byrd2010,byrd2012-a} involving
rubidium using the Karlshruhe def2-nZVPP basis sets
\cite{weigend2003,weigend2005} (n=T,Q zeta quality basis sets with two extra
$spdf$ correlation polarization functions) have shown good experimental
agreement for both dissociation energies and bond lengths of the Rb$_2$ diatom.
For rubidium these basis sets use the small-core ECP28MWB \cite{leininger1996}
Stuttgart pseudopotential, which removes the argon core electrons from the
calculation while leaving the 4s4p5s electrons free for use in further
correlation calculations.  The addition of even tempered $spdf$ diffuse
functions to these basis sets was done to better describe the anion, while the
addition of these diffuse functions has also shown to improve molecular
properties \cite{furche2010,byrd2011} as well.  This aug-Def2-nZVPP basis set was
used for rubidium in all riv electronic structure calculations in this work.  
To best describe the OH bond, the optimized aug-cc-pVnZ valence
\cite{kendall1992} and aug-cc-wCVnZ weighted core-valence \cite{peterson2002}
correlation basis sets were used for hydrogen and oxygen respectively.

Molecular structures were optimized using the CCSD(T)/riv level of
theory using the quadruple zeta (QZ) quality basis sets \footnote{It should be noted that bond
lengths computed using MP2 theory differ only by a few m\AA~ from CCSD(T)
calculations using the same basis set, at a much cheaper computational cost}.
Frequency calculations at the riv CCSD(T) level of theory using the QZ
basis sets were performed for each optimized structure to identify whether the
structure was a transition state or a local minimum of the potential energy
surface.  The final ground state structure of the RbOH$^-$ ion is found to be linear,
consistent with the ground state structure of the neutral molecule
\cite{lee2003,lara2007}.  Additionally the conformers OHRb$^-$ and ORbH$^-$ were
also investigated and found to be transition states.
Vibrational harmonic zero-point energy (ZPE) corrections were computed for the
final structures at the CCSD(T)/riv level of theory using the QZ basis set.
Computed bond lengths and ZPE corrections are listed in Table \ref{geomtable}.

\begin{table}[t]
\begin{ruledtabular}
\begin{tabular}{lrrr}
method & space & Rb$+$OH$^-$ & Rb$+$OH\\
\hline
        CCSD(T)/TZ & riv    &  75.40 & 129.25\\
         CCSD(T)/QZ & riv   &  74.63 & 130.47\\
\hline
CCSD(T)/Extrap. & riv       &  74.07 & 131.36\\
$\Delta$CCSD(T)/ANO & riiv  &   0.25 &   0.21\\
Total Energy & -            &  74.20 & 131.54
\end{tabular}
\end{ruledtabular}
\caption{\label{rbohtable}Breakdown of the contributions of various levels of
theory to the Rb-O bond energy (in $10^{-3}$ a.u.) for both neutral and charged RbOH.}
\end{table}

Correlation calculations involving the riiv electrons of rubidium involve
electrons dropped by the MWB pseudopotential.  Because of this we perform the
riiv correlation calculations all electron using the Roos atomic natural orbital
(ANO) basis set \cite{roos2004}, which was chosen for its availability for all
atoms present and its noted consistency \cite{hill2012}.  Prior to use, the
basis set was completely uncontracted so as to be as flexible as possible in
subsequent correlation calculations. Scalar relativistic effects were accounted
for by adding in the one-electron second-order Douglas-Kroll-Hess
\cite{douglas1974,wolf2002} contribution.  For rv and riv calculations, which
use the Def2 basis sets for rubidium, the small core MWB 
family of pseudopotentials have been shown \cite{leininger1996} to accurately
account for the relativistic contributions to the bond length and dissociation
energy.  

\begin{figure}[t]
\includegraphics[width=\columnwidth]{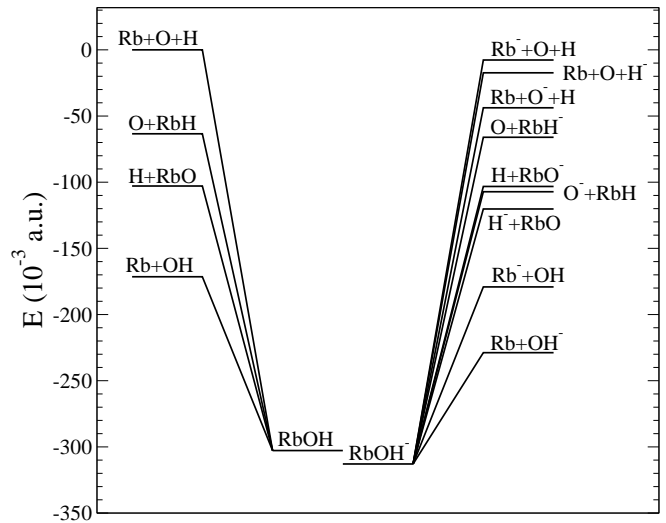}
\caption{\label{disslimit}Dissociation limits of rubidium hydroxyl (left) and its
anion (right) up to the atomization limit.  Energies are computed at the
CCSD(T)/CBS level of theory.}
\end{figure}

The complete basis set limit (CBS) of the various contributions to the total
energy was estimated using the two point linear extrapolation formula of
Helgaker {\it et al.} \cite{helgaker1997},
\begin{equation}\label{cbs}
E_{\rm CBS}({\rm method})=\frac{n^3 E_n - (n-1)^3 E_{n-1}}{n^3-(n-1)^3}.
\end{equation}
This extrapolation scheme was chosen over other more optimized schemes due to
the spread of basis sets and correlation spaces used here.
The final interaction energy is computed from the various contributions by the
following formula
\begin{equation}
E_{\rm int} = 
E_{\rm CBS}({\rm CCSD(T)/riv})  + 
E({\rm \Delta CCSD(T)/riiv}),
\end{equation}
where $E({\rm CCSD(T)/riv})$ is the total CCSD(T)/riv energy and
\begin{eqnarray}
E({\rm \Delta CCSD(T)/riiv}) & = &
E({\rm CCSD(T)/ANO/riiv}) \nonumber \\
&&-  E({\rm CCSD(T)/ANO/riv})
\end{eqnarray} 
is the riiv contribution.  Higher order triples contributions beyond the CCSD(T)
level of theory were estimated by performing CCSDT/rv calculations using QZ
quality basis sets.  Effects of connected quadruple excitations, known
\cite{martin2001} to be important for OH$^-$, were estimated using CCSDT(Q)/rb
with the triple zeta (TZ) quality basis sets.  It was found that the contribution of these
higher order terms to the final {electron affinity (EA) are small ($<5\times10^{-5}$ a.u.)} 
due to
cancellation.  While the riiv correlation contribution is similarly small for the
EA of RbOH at the equilibrium geometry, it becomes more significant for much
shorter Rb-O bond lengths (further discussed below).  In Table \ref{rbohtable},
the contributions of each of these corrections to the Rb-O bond energy are
listed.  Molecular bond lengths, electron affinities (EA) and atomization
energies (AE) (including the harmonic ZPE correction) are reported in Table
\ref{geomtable}.  The excellent agreement with available experimental bond
lengths, harmonic frequencies and electron affinities leads us to expect
comparable accuracy for the RbOH complex.

\begin{figure}[t]
\includegraphics[width=\columnwidth]{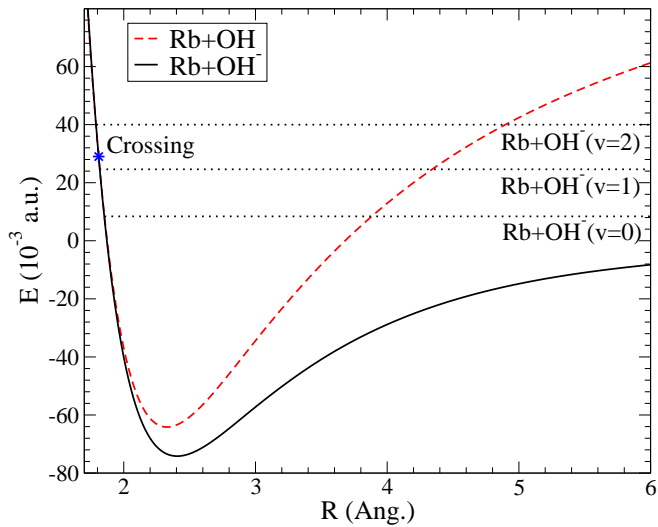}
\caption{\label{pescurve}Potential energy curve of Rb+OH and its anion computed
along the C$_{\infty v}$ axis at the CCSD(T)/CBS level of theory.  The OH bond
length is held fixed over the entire curve for simplicity.  Also shown are the
energy asymptotes for various OH$^-$ vibrational levels.}
\end{figure}

\section{Computational Results and Discussion}

\begin{figure}[t]
\includegraphics[width=\columnwidth]{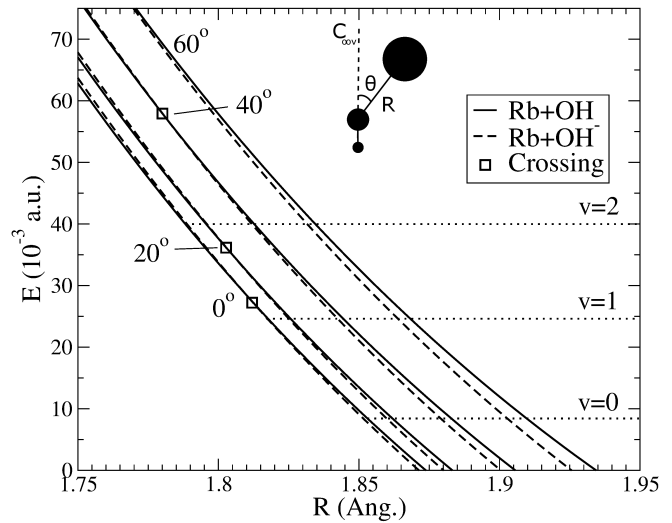}
\caption{\label{crossings}Inner wall potential energy curve of Rb+OH and its anion computed
for various collisional angles at the CCSD(T)/CBS level of theory with the OH bond
length relaxed at each point of the curve. 
{The sketch defines the angle $\theta$ of the Rb (largest circle)
approach to OH$^-$ centered on the oxygen (with O larger than H) from the C$_{\infty v}$ axis.}}
\end{figure}

The EA and similar geometric structure of rubidium hydroxyl and its anion, along
with the large difference in neutral and anion dissociation limits illustrated
in Fig. \ref{disslimit}, does not suggest immediately a charge loss pathway.  In
fact, at $300$ Kelvin no other dissociation channels are energetically accessible.
This is clearly illustrated in Fig. \ref{pescurve} where the minimum energy
dissociation path of Rb for both RbOH and its anion are computed at the fixed OH
bond CCSD(T)/riv/CBS level of theory.  It can be seen that the neutral and anion
curves do not cross at any point along the incoming path.  However it should be
noticed that the inner wall of these curves become nearly degenerate at this
level of theory.  We examine the inner wall more closely, by relaxing the OH
bond at each Rb-O distance using MP2 gradients (as noted previously, MP2 bond
lengths are very close to CCSD(T) bond lengths), and find that indeed the
neutral and anion curves cross at r(Rb-O)$\sim1.81$\AA~ with a barrier height of
{$V_c(0)\sim3.0\times10^{-2}$} a.u. above the Rb+OH$^-$ dissociation limit, as
illustrated in Fig. \ref{crossings}.  This crossing
energy $V_c(\theta)$ also includes the CCSD(T)/riiv correction which provides 
{ $\sim4\times10^{-4}$} a.u. to the final barrier height.  
This crossing is energetically accessible
if the internal rotational and vibrational energy of OH$^-$ is taken into
account.  In fact, it is well known that producing rotationally and
vibrationally cool OH$^-$ is difficult experimentally \cite{rosenbaum1986}.

The height of $V_c(\theta)$ for geometries other than the linearly minimum energy
approach was also investigated.  It was found that for small angle approaches,
relative to the equilibrium geometry, the crossing remains relatively flat,
while for angles greater than $40$ degrees the barrier rapidly increases in
height until it is completely energetically inaccessible (see Fig.
\ref{crossings}).  

To evaluate the associative detachment rate coefficient, while accounting for
the dependence with the angle of approach, we assume that OH$^-$, in its 
internal ro-vibrational state $(v,J)$, is rotating sufficiently fast during its encounter with Rb. We 
then average the angular dependence of $V_c(\theta)$ to obtain an effective 
angular phase space $\rho_c(v,J,\varepsilon)$ where the curve crossing is accessible
for a given collision energy $\varepsilon$,
\begin{equation}\label{rhoc}
   \rho_c(v,J,\varepsilon) = \frac{1}{2} 
   \int_0^\pi d(\cos\theta )
   \Theta\left[(\varepsilon-V_c(\theta)+T(v,J)\right] .
\end{equation}
Here, the prefactor $\frac{1}{2}$ arises from the azimuthal 
angle integration, $\Theta(\cdots)$ is a Heaviside step function representing the
height of the crossing as a function of the collision angle and for a given
collision energy $\varepsilon$, and $T(v,J)=G(v)+F_v(J)$ is the internal 
rotation-vibration energy of the
OH$^-$ fragment in its $v$'th vibrational and $J$'th rotational state,
which we take to follow a Dunham series \cite{dunham1932}.
The integral over $\theta$ involving $V_c(\theta)$ is performed numerically,
with representative values presented in Table \ref{rhoctable} for a few $J$'s of
$v=0,\dots,3$ at $\varepsilon / k_B = 300$ K ($k_B$ is the Boltzmann constant). 
We find that $\rho_c(v,J,\varepsilon)$
is negligible for $v=0$ and $v=1$ for low $J$'s, and becomes more significant, 
reaching the range of 10-20\% for $v\geq 2$ in this table.

Using this approximation, we estimate the total cross section for 
associative detachment $\sigma_{\rm tot}(v,J,\varepsilon)$
by multiplying the Langevin cross section $\sigma_L(\varepsilon)$
for entering the inner region of the Rb+OH$^-$ curve (where the
process can take place with assumed unit probability) by the fraction of 
angular phase space $\rho_c(v,J,\varepsilon)$ allowing the process 
({\it i.e.} when the curve crossing is accessible)
\begin{equation}\label{sigmatot}
   \sigma_{\rm tot}(v,J,\varepsilon)=
   \sigma_L(\varepsilon)\rho_c(v,J,\varepsilon) .
\end{equation}
Here, $\sigma_L(\varepsilon)$ depends on the static dipole polarizability
$\alpha_d$ of the neutral monomer (with $\alpha_d = 318.6$ for Rb
\cite{derevianko1999}),
but not on the inner part of the potential curve \cite{levine1987,cote2000}
\begin{equation}
    \sigma_L(\varepsilon)=\pi\sqrt{\frac{2\alpha_d}{\varepsilon}} .
\end{equation}

\begin{table}[b] 
\caption{\label{rhoctable}Representative values of the
accessible angular space, Eq.(\ref{rhoc}), for various vibrational and
rotational states for a collisional energy of $300$ Kelvin.}
\begin{ruledtabular}
\begin{tabular}{rrrrr}
 & \multicolumn{4}{c}{$\rho_c(v,J,300K)\times 100$} \\
$J$ & $v=0$ & $v=1$ & $v=2$ & $v=3$\\
\hline
0  & 0.00 & 0.00 & 5.38 & 11.00 \\
5  & 0.00 & 0.00 & 6.56 & 12.07 \\
10 & 0.00 & 3.16 & 9.53 & 14.80 \\
15 & 0.89 & 8.11 & 14.00 & 18.92 \\
\end{tabular}
\end{ruledtabular}
\end{table}

\begin{figure}
\includegraphics[width=\columnwidth]{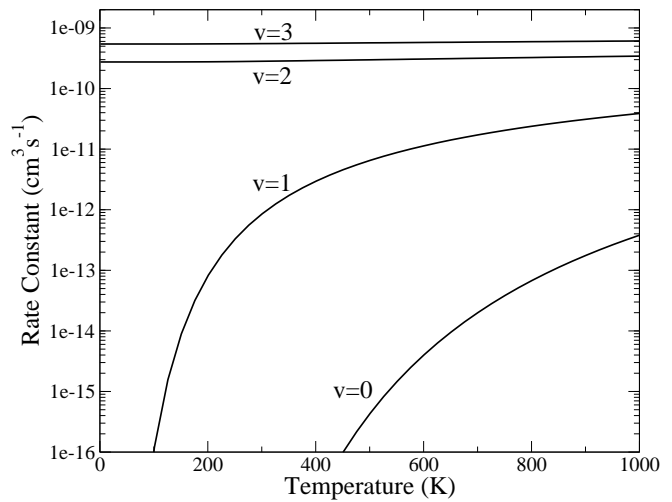}
\caption{\label{ratecurve}Ion capture rate coefficient as a function of the
collisional temperature of the OH$^-$ complex for the first four vibrational
levels.}
\end{figure}

\begin{figure}
\includegraphics[width=\columnwidth]{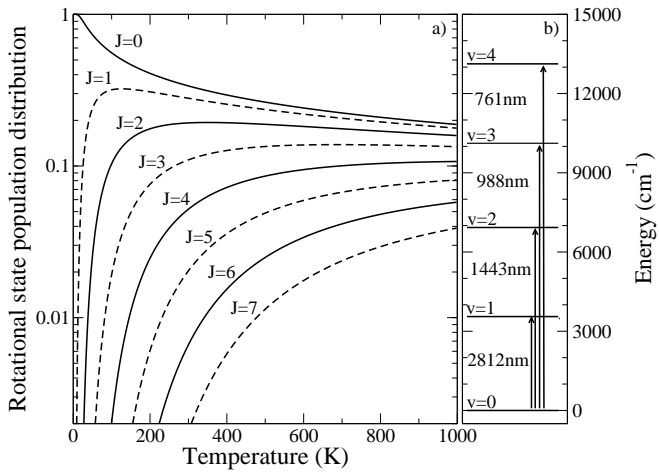}
\caption{\label{jpop}a$)$ Rotational state population distribution as a function
of temperature for the $v=0$ state of OH$^-$.  Effects on the rotational
distribution as a function of vibrational level are quantitative only and vary
on the order of $3\%$.  b$)$ Rotationless vibrational level energies,
$T(v,0)$, for OH$^-$ and { transition wavelengths} from $v=0$ to 
various excited vibrational levels.}
\end{figure}

{
A thermal rate constant $k_{\rm ad}(v,J)$ for the associative detachment for OH$^-$
initially in a specific $(v,J)$ state is obtained by 
averaging $v \sigma_{\rm tot}$ 
over a Maxwell distribution of velocity $v$ characterized by a (translational) temperature  
$T$, namely
\begin{equation}\label{klvj}
   k_{\rm ad}(v,J) = \sqrt{\frac{2k_B T}{\pi\mu}} 
   \int_0^\infty d x\; x \; e^{-x} \sigma_{\rm tot}(v,J,x k_B T) ,
\end{equation}
where $x=\varepsilon/k_B T$, and $\mu$ is the reduced mass of the colliding partners
(here Rb and OH$^-$). 
}
The distribution of rotational states $J$ is assumed to be
thermalized against the collision energy, which gives the vibrational state
rate constant 
{
\begin{equation}\label{klv}
k_{\rm ad}(v)=\frac{1}{Q_{\rm rot}}
\sum_J^{J_{\rm max}} k_{\rm ad}(v,J) \exp\left(-\frac{F_v(J)}{k_B T}\right),
\end{equation}
}
with the rotational partition function given by
{
\begin{equation}\label{qrot}
Q_{\rm rot}=\sum_J^{J_{\rm max}} \exp\left(-\frac{F_v(J)}{k_B T}\right) .
\end{equation}}
Here $J_{\rm max}$ is the maximum rotational state taken in the series.  To
evaluate Eq.(\ref{klv}) we use the spectroscopic constants of Rosenbaum {\it et
al.} \cite{rosenbaum1986} in $T(v,J)$
\footnote{$B_e=19.12087$ cm$^{-1}$, $\alpha_e=0.77167$ cm$^{-1}$, $\omega_e=3738.44$ cm$^{-1}$ and
$\omega_e x_e=91.42$ cm$^{-1}$ \cite{rosenbaum1986}}
and choose $J_{\rm max}$ such that the
thermodynamic contribution of that rotational state is negligible (see Fig.
\ref{jpop} a)).  A value of
$J_{\rm max}=15$ was found to be more than adequate to converge the sums in
Eqs.(\ref{klv})-(\ref{qrot}) even for very high temperatures.  
{ The rate constant (\ref{klv}) was}
evaluated numerically for the first three
vibrational levels of OH$^-$ as a function of the collisional temperature, the
results of which are plotted in Fig. \ref{ratecurve}.  
{ The rate constant for $v=0$ and
$v=1$ is found to be much lower than a recent experimental value
\cite{deiglmayr2012}, but becomes comparable for $v=2$, as expected
considering the energetics of the collision. }
It should be noted that for $v=2$ and higher, the incoming collisional
energy is above the curve crossing threshold, and so the rate coefficient is a nearly constant as
expected.  

\begin{table}[b]
\caption{\label{ratetable}Spontanious emission rotational rate coefficients,
Eq.(\ref{spontem}), for OH$^-$ in various starting rotational and vibrational levels.
Units are in inverse seconds and $[n]$ represents $\times 10^n$.}
\begin{ruledtabular}
\begin{tabular}{lllllll}
 & \multicolumn{6}{c}{$A_{v'\rightarrow v}(J')$ (s$^{-1}$)}\\
$J'$ & 
\multicolumn{1}{l}{$1\rightarrow 0$} &
\multicolumn{1}{l}{$2\rightarrow 0$} &
\multicolumn{1}{l}{$2\rightarrow 1$} &
\multicolumn{1}{l}{$3\rightarrow 0$} &
\multicolumn{1}{l}{$3\rightarrow 1$} &
\multicolumn{1}{l}{$3\rightarrow 2$} \\
\hline
0 & 6.87[1] & 5.16[1] & 1.22[2] & 2.19[1] & 1.36[0] & 1.75[2]\\
1 & 2.36[2] & 1.78[0] & 4.18[2] & 7.54[1] & 4.68[0] & 6.00[2]\\
2 & 4.34[2] & 3.26[0] & 7.68[2] & 1.38[0] & 8.58[0] & 1.10[3]\\
3 & 6.31[2] & 4.72[0] & 1.12[3] & 2.00[0] & 1.24[1] & 1.61[3]\\
4 & 8.30[2] & 6.20[0] & 1.47[3] & 2.63[0] & 1.64[1] & 2.12[3]\\
5 & 1.03[3] & 7.70[0] & 1.83[3] & 3.27[0] & 2.03[1] & 2.64[3]\\
6 & 1.24[3] & 9.23[0] & 2.21[3] & 3.91[0] & 2.44[1] & 3.19[3]\\
7 & 1.46[3] & 1.08[1] & 2.60[3] & 4.58[0] & 2.86[1] & 3.76[3]\\
8 & 1.69[3] & 1.24[1] & 3.01[3] & 5.26[0] & 3.29[1] & 4.36[3]\\
9 & 1.92[3] & 1.41[1] & 3.43[3] & 5.97[0] & 3.74[1] & 4.98[3]\\
10 & 2.17[3] & 1.59[1] & 3.88[3] & 6.71[0] & 4.20[1] & 5.65[3]\\
11 & 2.43[3] & 1.77[1] & 4.36[3] & 7.47[0] & 4.69[1] & 6.36[3]\\
12 & 2.70[3] & 1.96[1] & 4.86[3] & 8.27[0] & 5.20[1] & 7.11[3]\\
13 & 3.00[3] & 2.16[1] & 5.40[3] & 9.10[0] & 5.74[1] & 7.91[3]\\
14 & 3.30[3] & 2.36[1] & 5.97[3] & 9.97[0] & 6.30[1] & 8.77[3]\\
15 & 3.63[3] & 2.58[1] & 6.57[3] & 1.09[1] & 6.89[1] & 9.69[3]
\end{tabular}
\end{ruledtabular}
\end{table}

As these collisions involve rotationally and vibrationally excited states of a
polar molecule, it is important to characterize the lifetime of these states in
the absence of collisions.  To do this we have computed the OH$^-$ potential
energy curve near the equilibrium geometry at the $E_{\rm CBS}$(CCSD(T)/riv)
level of theory as well as the static dipole moment ${\cal D}(R)$ along this
curve at the CCSD(T)/rv level of theory using the aug-cc-pVQZ basis set.  We
find that the resulting dipole moment function is in good agreement 
\footnote{The dipole moment function and near equilibrium potential energy curve
are plotted in the supplemental material.}
with the
original work of Werner {\it et al.} \cite{werner1983}. 
The spontaneous emission rate for a given initial
rotational and vibrational state $J',v'$ radiating to all possible lower energy
rotational states is (in atomic units \footnote{The conversion from atomic units
to inverse seconds used in this work is $1/(2.419\times10^{-17}s)$}) given by
\begin{multline}\label{spontem}
{A_{v'\rightarrow v}(J')} = 
\sum_{J} \frac{4}{3} \alpha^3 (T(v',J')-T(v,J))^3\times\\
\langle v';J'M'\Omega'|{\cal D}(R)|JM\Omega;v\rangle,
\end{multline}
where $|v\rangle$ is the vibrational wave function for the $v$th level and
$|JM\Omega\rangle$ is the rigid-rotor wavefunction \cite{browncarrington}.  We
have tabulated our results for $A_{J',v'\rightarrow v}$ in Table \ref{ratetable}
for all $J'$ up to $J_{max}$, for the various vibrational transitions for $v=0$
through $v=3$ (see Fig. \ref{jpop} b)).  Given an inital vibrational state, the temperature dependent
lifetime averaged over rotational level is
{
\begin{equation}\label{tlifetime}
   \tau_{v'} = \frac{1}{Q_{rot}}\sum_v \sum_{J}^{J_{max}}
   \frac{\exp\left(-F_v(J)/k_B T\right)} {A_{v'\rightarrow v}(J')}. 
\end{equation} }
In Table \ref{lifetimet} the lifetimes for the first $3$ excited vibrational
levels are listed at several representative temperatures.

\begin{table}[t]
\caption{\label{lifetimet}Temperature dependent vibrational state lifetimes
averaged over inital rotational states.  Units are in kelvin and miliseconds.}
\begin{ruledtabular}
\begin{tabular}{dddd}
\multicolumn{1}{c}{temperature} &
\multicolumn{1}{c}{$v'=1$} &
\multicolumn{1}{c}{$v'=2$} &
\multicolumn{1}{c}{$v'=3$} \\
\hline
10  & 14.5 & 8.17 & 5.67 \\
100 & 9.94 & 5.60 & 3.88 \\
300 & 6.98 & 3.93 & 2.72 \\
600 & 5.44 & 3.06 & 2.12
\end{tabular}
\end{ruledtabular}
\end{table}

\section{Conclusions}

We have computed CCSD(T) {\it ab initio} potential energy curves for the Rb+OH
and Rb+OH$^-$ systems, and found a neutral-ion curve crossing along the inner
wall of the potential energy curve for collinear geometries of RbOH.  Further
investigation of the potential energy surface shows that this crossing is highly
dependent on collisional angle and is accessible to reasonable scattering
energies only for angles near the collinear geometry. Furthermore, this crossing
lies above the OH$^-$ $(v=0)$ collisional threshold, and so is expected to have a
negligible contribution to the long-term co-trapping of rubidium and hydroxide.
Using the Langevin capture cross-section, we evaluate the associative
detachment rate for the first few vibrationally excited states of hydroxide
colliding with rubidium and find an appreciable { rate coefficient} $>2\times 10^{-9}$
cm$^3$s$^{-1}$ for hydroxide vibrational levels $v\ge 2$.  Lifetimes for these
vibrationally excited hydroxide molecules are computed, and found to be on the
order of $5$ ms.  With transitions between $v=0$ and $v\ge 2$ (see Fig.
\ref{jpop} b)) in the near infrared, it is possible to directly control access to the associative
recombination pathway that forms neutral RbOH.


\section{Acknowledgments}

We would like to thank M. Weidem\"uller for many helpful conversations during
the course of this work.  { The work of J.N.B. was partially supported by the 
US Department of Energy, Office of Basic Energy Sciences, and the work of R.C.
by the National Science Foundation Grant No. PHY 1101254.}


%

\end{document}